\documentclass[11pt,a4paper]{quantumarticle}
\pdfoutput=1

\usepackage{amsmath}
\usepackage{amssymb}
\usepackage[english]{babel}
\usepackage{booktabs}
\usepackage{cite}
\usepackage{graphicx}
\usepackage{hyperref}
\usepackage{listings}
\usepackage{multirow}
\usepackage{siunitx}
\usepackage{color}
\usepackage{xcolor}

\definecolor{bgcolor}{HTML}{F4EFDE}
\definecolor{kwcolor}{HTML}{385bb1}

\begin{document}

\title{Qade: Solving Differential Equations on Quantum Annealers}

\author{Juan Carlos Criado}
\email{juan.c.criado@durham.ac.uk}
\affiliation{Institute for Particle Physics Phenomenology, Durham University, Durham DH1 3LE, UK}
\affiliation{Department of Physics,  Durham University, Durham DH1 3LE, UK}

\author{Michael Spannowsky}
\email{michael.spannowsky@durham.ac.uk}
\affiliation{Institute for Particle Physics Phenomenology, Durham University, Durham DH1 3LE, UK}
\affiliation{Department of Physics, Durham University, Durham DH1 3LE, UK}

\maketitle

\begin{abstract}
 We present a general method, called Qade, for solving differential equations using a quantum annealer. The solution is obtained as a linear combination of a set of basis functions. On current devices, Qade can solve systems of coupled partial differential equations that depend linearly on the solution and its derivatives, with non-linear variable coefficients and arbitrary inhomogeneous terms. We test the method with several examples and find that state-of-the-art quantum annealers can find the solution accurately for problems requiring a small enough function basis. We provide a Python package implementing the method at \href{https://gitlab.com/jccriado/qade}{gitlab.com/jccriado/qade}.
\end{abstract}


\section{Introduction\label{sec:intro}}

The development of methods for the solution of differential equations is fundamental for the mathematical modelling of most systems in nature. Methods based on machine learning techniques, first proposed in Refs.~\cite{LEE1990110, MEADE19941, MEADE199419, Lagaris:1997at}, have gained increasing attention for their versatility in recent years~\cite{raissi2017physics, raissi2017physicsII, RAISSI2019686, Han8505, magill2018neural,  Piscopo:2019txs, dockhorn2019discussion, REGAZZONI2019108852, chen2019neural, shen2020deep, 6658964, RUDD2015277, Sirignano_2018}. At their core, these methods reformulate the task of solving differential equations as an optimization problem, for which the existing machine learning frameworks are designed~\cite{DBLP:journals/corr/abs-1907-04502, DBLP:journals/corr/abs-1909-11544, hennigh2020nvidia,Chen2020, hartmann2020neural, JIN2021109951, li2020fourier, lau2020oden, guidetti2021dnnsolve, Araz:2021hpx}.

While classical machine learning approaches have shown to be able to solve differential equations by minimising the network's loss function, a common issue in classical optimization algorithms is the difficulty in escaping deep local minima. Quantum computing provides an immediate solution: the quantum tunnelling mechanism allows to jump between local minima separated by large energy barriers~\cite{Abel:2020ebj}. In this way, the global minimum of non-convex functions can be found reliably~\cite{Abel:2021fpn,Abel:2022lqr}.

Quantum annealing devices~\cite{finilla94a, kadowaki98a, brooke99a, dickson13a, lanting14a, albash15a, albash16a, boixo16a, chancellor16b, Benedetti16a, Muthukrishnan2016, cervera, LantingAQC2017} are particularly well-suited for optimization tasks, as the computation they perform is directly the minimization of their Hamiltonian, which the user can specify. Thus, to unleash the prowess of a quantum annealer for optimization tasks, one needs to encode the problem as an Ising model. A general method for approximately encoding arbitrary target functions with the compact domain as the Hamiltonian of a quantum annealer has been introduced in Ref.~\cite{Abel:2022lqr}. This can be combined with methods that use a coarse-grained approximation and iteratively improve the precision of the solution, as proposed in Ref.~\cite{Zlokapa:2019lvv}, to reduce the necessary number of qubits and connections between them.

In this work, we apply these quantum optimization techniques to the solution of differential equations in the machine learning-oriented formulation described in Refs.~\cite{Piscopo:2019txs, Araz:2021hpx}. Previous implementations of other methods in quantum computers have been applied successfully to solving differential equations in Refs.~\cite{PhysRevA.101.010301, zanger2021quantum, PhysRevA.99.052355, kyriienko2021solving}. The main advantage of our method, which we call Qade, is its generality: it does not make assumptions about the equations or boundary conditions beyond linearity. This means that systems of coupled linear partial differential equations of any order, with variable coefficients and arbitrary inhomogeneous terms, can be handled by this approach.

We provide a Python package implementing it in full generality with the method. This package contains the tools for obtaining the quantum annealing formulation of linear differential equations. Furthermore, users with access to the cloud interface to the D-Wave quantum annealers can also perform the necessary annealing runs and decode the results into the final solution.

The rest of this paper is organized as follows. In Section~\ref{sec:quantum-annealing}, we briefly introduce the quantum annealing framework. In Section~\ref{sec:method}, we present the Qade method, which reformulates the task of solving differential equations as a problem directly solvable by quantum annealers. Examples of application of Qade to 3 differential equations are shown in Section~\ref{sec:examples}. We summarize our conclusions in Section~\ref{sec:conclusions}. Finally, the accompanying Python package is introduced in Section~\ref{sec:qade}.

\section{Quantum annealing\label{sec:quantum-annealing}}

In the quantum annealing paradigm, computations are encoded as finding the ground state of an Ising model Hamiltonian
\begin{equation}
  H(\sigma) = \sum_{ij} \sigma_i J_{ij} \sigma_j  + \sum_i h_i \sigma_i,
\end{equation}
for a collection of $N$ spin variables $\sigma_i = \pm 1$. To perform a quantum annealing calculation, one must then find a way to reduce the problem at hand to the minimization of $H$. In Section~\ref{sec:method}, we describe how to obtain the $J_{ij}$ and $h_i$ parameters corresponding to any system of linear partial differential equations.

We now review how $H$ is minimized in a quantum annealing device. Internally, the device has access to a quantum system that it partially controlles. The system is described by a Hilbert space constructed as the tensor product of $N$ 1-qubit spaces $\mathbb{C}^2$ and a Hamiltonian
\begin{multline}
  H_{\text{quantum}} =
 A(s) \sum_i \sigma_{ix} \\
  + B(s) \left[\sum_{ij} \sigma_{iz} J_{ij} \sigma_{jz}  + \sum_i h_i \sigma_{iz} \right],
\end{multline}
with $\sigma_{ia}$ the $a$th Pauli matrix applied to the $i$th qubit, and $A(1) = B(0) = 0$. The annealer can set the values of $J_{ij}$ and $h_i$, prepare the system in the ground state of $H_{\text{quantum}}$ at $s=0$, change the value of $s$ continuously, and measure the observable $\bigotimes_i \sigma_{iz}$ at the end of the annealing process, when $s=1$.

The dependence $s(t)$ of the $s$ parameter with time $t$ is referred to as an schedule. A typical schedule is a monotonically increasing from function $s=0$ at the initial time to $s=1$ at the final time, which, depending on the application, can vary between a few $\si{\micro s}$ to about $\SI{1}{ms}$. A pause in the increase of $s(t)$ for some time or an increase in its slope towards the end of the run is commonly used. When an appropriate schedule is selected, the final measurement of the annealer is expected to return the ground state of $H$.

The most expensive part of the computation is the preparation of Hamiltonian with the specified parameters $J_{ij}$ and $h_i$. Once this is done, the annealing process is usually run several times, and the final state with the minimal solution is selected to reduce noise.

\section{Method\label{sec:method}}

We now present Qade, our quantum annealing-based method for solving differential equations. We denote the equations to be solved as
\begin{equation}
  \left. E_i(x)[f] \right|_{x \in \mathcal{X}_i} = 0,
  \label{eq:equations}
\end{equation}
for a function $f : \mathbb{R}^{n_{\text{in}}} \to \mathbb{R}^{n_{\text{out}}}$, where the $E_i(x)$ are local functionals of $f$ (i.e., they only depend on the value of $f$ and its derivatives at $x$), and the $\mathcal{X}_i$ are the domains in which the equations must be satisfied. Initial and boundary conditions can be viewed as a particular case of these equations, in which $\mathcal{X}_i$ is the initial or boundary set of $x$ values. We impose that all the equations are linear functions of $f$ and its derivatives:
\begin{equation}
  E_i(x)[f] = \sum_{kn} C^{(k)}_{in}(x) \cdot (\partial^k f_n(x)) + B_i(x),
\end{equation}
where the $B_i(x)$ are the inhomogeneous terms, while, the $C^{(k)}_{in}(x)$ are the variable coefficients of the derivatives, and $\partial^k f_n$ is a vector containing all the partial derivatives of order $k$ of $f_n$.

As explained in Section~\ref{sec:quantum-annealing}, to solve Eq.~\ref{eq:equations} in a quantum annealer, it has to be encoded as the ground state of an Ising model Hamiltonian. We first reformulate it as a minimization problem. Following the machine learning-oriented methods described in Refs.~\cite{Piscopo:2019txs, Araz:2021hpx}, we discretize the domains into finite subsets of sample points $X_i \subset \mathcal{X}_i$, and define the loss function
\begin{equation}
 L[f] = \sum_{i} \sum_{x \in X_i} {\left(E_i(x)[f]\right)}^2.
\end{equation}
The global minimum $L[f_{\text{sol}}] = 0$ is attained if and only if all the equations are satisfied at all the sample points in the $X_i$ sets. Now, we parametrize the function $f$ as a linear combination of a finite set of ``basis'' functions $\Phi_m$, as
\begin{equation}
 f_n(x) = \sum_m w_{nm} \Phi_m(x).
 \label{eq:basis-functions}
\end{equation}
Then, the equations can be written as linear functions of a finite set of parameters, the weights $w_{nm}$:
\begin{align}
  E_i(x, w) &= \sum_{nm} H_{in}(x)[\Phi_m] \; w_{nm} + B_i(x), \\
  H_{in}(x)[\Phi] &= \sum_k C^{(k)}_{in}(x) \cdot (\partial^k \Phi(x)),
\end{align}
and the $L$ becomes a quadratic function of them:
\begin{equation}
  L(w) = \sum_{nmpq} w_{nm} \, J_{nm,pq} \, w_{pq} + \sum_{nm} h_{nm} \, w_{nm},
  \label{eq:quadratic-loss}
\end{equation}
where
\begin{align}
  J_{nm, pq} &= \sum_i \sum_{x \in X_i} H_{in}(x)[\Phi_{m}] \; H_{ip}(x)[\Phi_{q}], \label{eq:quad-loss-J}
  \\
  h_{nm} &= 2 \sum_i \sum_{x \in X_i} H_{in}(x)[\Phi_m] B_i(x). \label{eq:quad-loss-h}
\end{align}

The final step in converting $L$ into an Ising model Hamiltonian is the binary encoding of each weight in terms of spin variables $\hat{w}^{(\alpha)}_{nm} = \pm 1$, as
\begin{equation}
  w_{nm} = c_{nm} + s_{nm} \sum_{\alpha = 1}^{n_{\text{spins}}} \frac{\hat{w}^{(\alpha)}_{nm}}{2^\alpha},
  \label{eq:binary-encoding}
\end{equation}
with the free parameter $c_{mn}$ and $s_{nm}$ being the center values of the $w_{nm}$, and the scales by which the can change within the encoding, respectively. Replacing this expression into Eq.~\eqref{eq:quadratic-loss}, we finally get the Ising model
\begin{multline}
  H(\hat{w}) := L(w)
  = \sum_{nmpq\alpha\beta}\hat{w}^{(\alpha)}_{nm} \, \hat{J}^{(\alpha\beta)}_{nm, pq} \,  \hat{w}^{(\beta)}_{pq} \\
  + \sum_{nm\alpha} \hat{h}^{(\alpha)}_{nm} \, \hat{w}^{(\alpha)}_{nm},
\end{multline}
where
\begin{align}
  \hat{J}^{(\alpha\beta)}_{nm, pq} &= 2^{-(\alpha + \beta)} s_{nm} \, s_{pq} \, J_{nm, pq}, \label{eq:ising-J}
  \\
  \hat{h}^{(\alpha)}_{nm} &= 2^{-\alpha} s_{nm} (h_{nm} + 2 c_{pq} \, J_{nm, pq}). \label{eq:ising-h}
\end{align}
The original problem can then be solved by minimizing $H$ in a quantum annealing device. The solution is recovered by decoding the weights using Eq.~\eqref{eq:binary-encoding}, and substituting them in Eq.~\eqref{eq:basis-functions}.

The size of an Ising model embedded in a current quantum annealer is limited, both in the allowed number of spins and the number of connections between them. This means that not many spins per weight $n_{\text{spin}}$ can be currently used, which implies that each weight can only be determined up to a low precision $2^{-n_{\text{spin}}}$ in a single quantum annealing run. To improve the accuracy of the results, we use a version of the iterative algorithm proposed in Ref.~\cite{Zlokapa:2019lvv}. In each iteration $I$, which we call an \emph{epoch}, the annealer is run for the model defined by setting the centres to the values of the weights obtained in the previous iteration, while all the $s_{nm}$ are scaled by a factor $0 < S \leq 1$:
\begin{equation}
 \begin{array}{c}
  c^I_{nm} = w^{I-1}_{nm}, \quad
  s^I_{nm} = S s^{I-1}_{nm}, \\
  I = 0,\ldots, n_{\text{epochs}}.
 \end{array}\label{eq:epoch-update}
\end{equation}
We remark that the use of $n_{\text{epochs}} > 1$ is due to the limited number of qubits and connections available in the physical device being used. In future annealers with a larger size, one might be able to set a larger $n_{\text{spins}}$ and $n_{\text{epochs}} = 1$, so that the solution is obtained in one annealing step, and one can take full advantage of the quantum computation.

The method we have presented contains several \emph{hyperparameters} that need to be adjusted to suitable values before application to a concrete problem. We collect them in Table~\ref{tab:hyperparameters}, together with examples of the typical values they might take to solve differential equations with current quantum annealing devices. To choose a correct set of values for the hyperparameters, one might use domain knowledge about the problem to be solved, such as which basis of functions is most suited or what is the typical size that the corresponding weights might have. When this knowledge is not available, the process involves some trial and error, using the value of loss function $L(w)$ as a measure of the goodness of the solution.

\begin{table*}
  \centering
  \begin{tabular}{crlc}
    \toprule
    & Example value & Description
    & Definition \\
    \midrule
    \multirow{2}*{annealing}
    & $n_{\text{reads}}= 200$ & number of reads
    & \multirow{2}*{Section~\ref{sec:quantum-annealing}} \\
    & $s(t) = t / (\SI{200}{\micro s})$ & quantum annealing schedule \\
    \midrule
    \multirow{3}*{encoding}
    & $n_{\text{spins}} = 3$ & number of spins per weight
    & \multirow{3}*{Eq.~\eqref{eq:binary-encoding}} \\
    & $c_{mn} = 0$ & (initial) central values of the weights  \\
    & $s_{mn} = 1$ & (initial) scales of the weights  \\
    \midrule
    \multirow{3}*{general}
    & $\Phi_m(x) = x^m$ & basis of functions
    & Eq.~\eqref{eq:basis-functions} \\
    & $n_{\text{epochs}} = 10$ & number of epochs in the iterative procedure
    & \multirow{2}*{Eq.~\eqref{eq:epoch-update}} \\
    & $S = 1/2$ & scale factor to update $s_{nm}$ in each epoch \\
    \bottomrule
  \end{tabular}
  \caption{Hyperparameters of the method presented in Section~\ref{sec:method}, with example values.\label{tab:hyperparameters}}
\end{table*}

\section{Examples\label{sec:examples}}

This section illustrates how to use the Qade method proposed in this paper to solve different kinds of differential equations. First, we solve equations whose solutions are known analytically to be able to compare them to the numerical results: the Laguerre equation, as an example of a single ordinary differential equation with variable coefficients; the wave equation, as an example of a partial differential equation; and an example of a first-order system of coupled differential equations. The code for these examples is available at \href{https://gitlab.com/jccriado/qade/-/tree/main/examples}{gitlab.com/jccriado/qade/-/tree/main/examples}.

\begin{figure}
  \centering
  \includegraphics[width=0.4\textwidth]{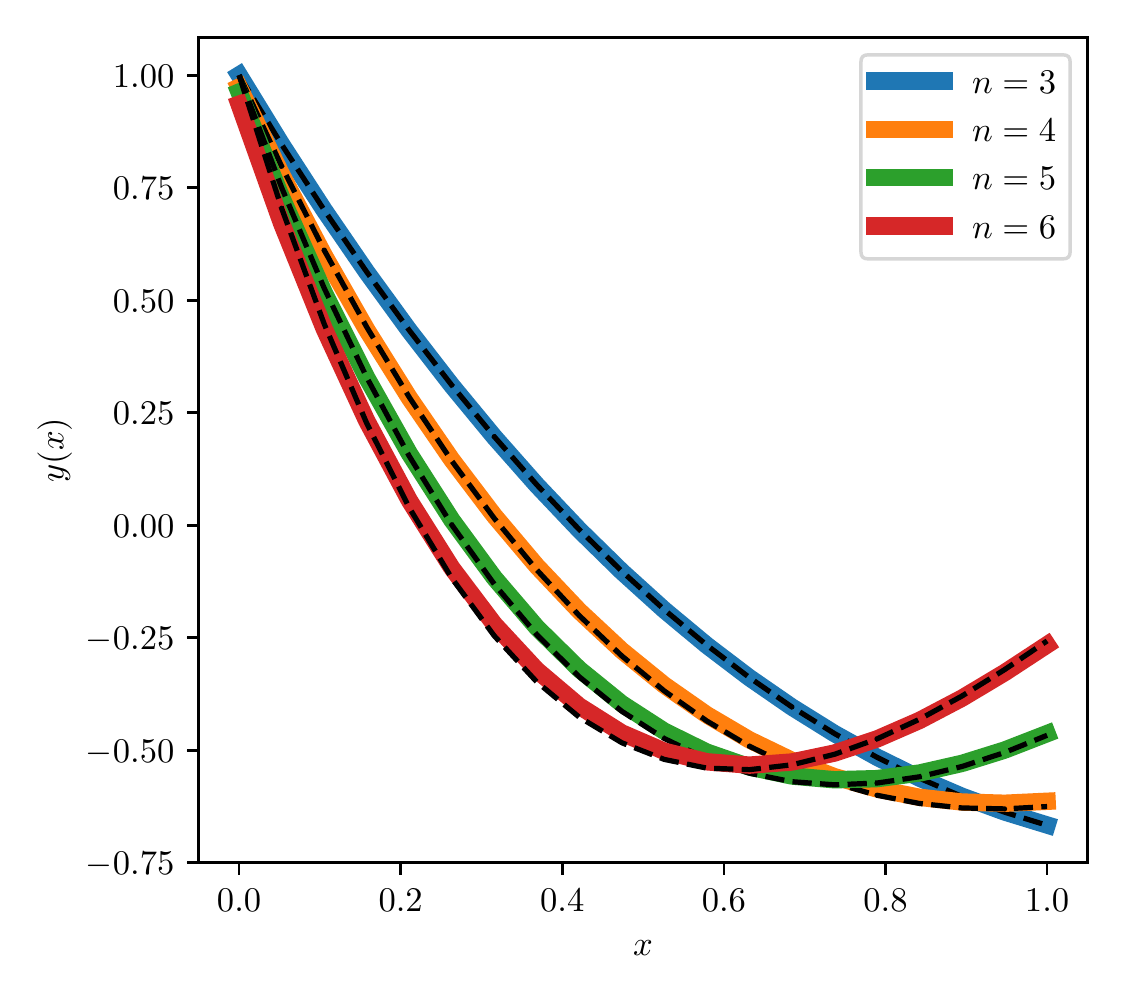}
  \includegraphics[width=0.4\textwidth]{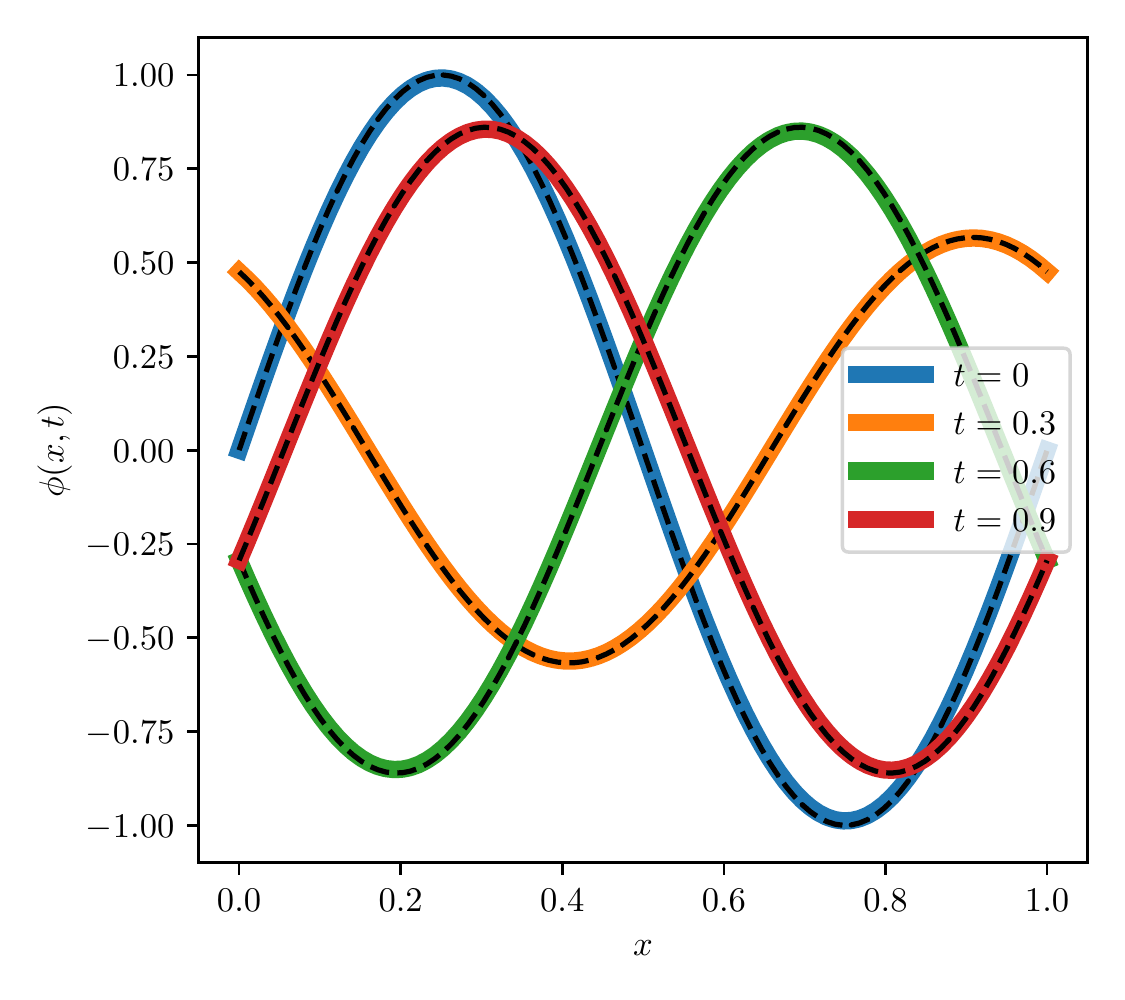}
  \includegraphics[width=0.4\textwidth]{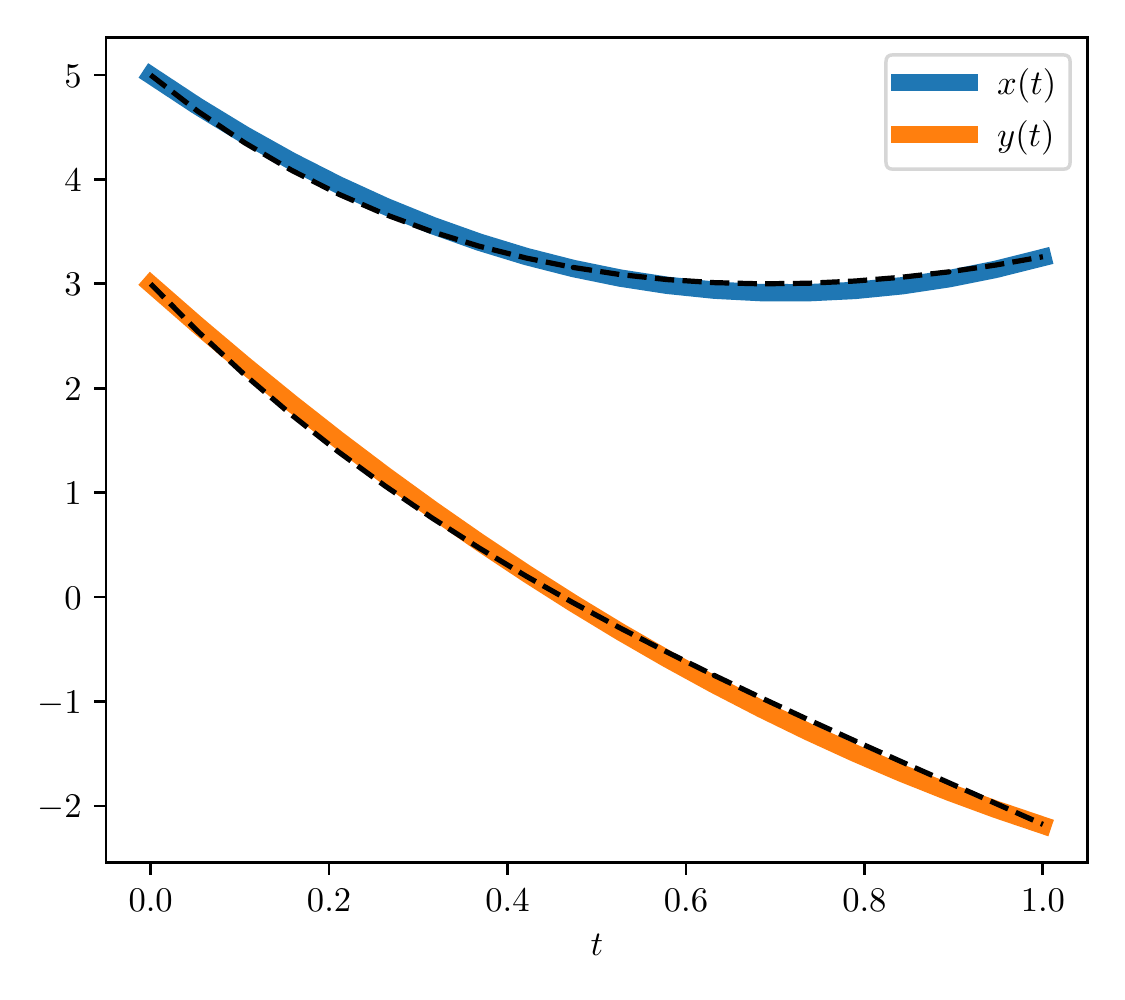}
  \caption{Solutions obtained with \texttt{Qade} to the Laguerre equation (top, see Section~\ref{sec:laguerre}), wave equation (center, see Section~\ref{sec:wave}) and a system of first-order coupled equations (bottom, see Section~\ref{sec:coupled}). The dashed lines show the analytical solutions.}
  \label{fig:examples}
\end{figure}

\subsection{Laguerre equation\label{sec:laguerre}}

The Laguerre equation is given by
\begin{equation}
  x y'' + (1 - x) y' + n y = 0.
\end{equation}
Its solutions are the Laguerre polynomials $L_n(x)$. We thus impose the boundary conditions
\begin{equation}
  y(0) = 1, \qquad y(1) = L_n(1),
\end{equation}
We look for a solution of the form $y = w_m x^m$, with $0 \leq m < 4$. We employ the Ising model formulation outlined in Section~\ref{sec:method} to find the weights $w_m$ using  D-Wave's \texttt{Advantage\_system4.1}. Since the weights $w_m$ are expected to grow for increasing $n$, we pick the scales $s_m = n - 2$. We also find that setting a high $n_{\text{reads}} = 500$ gives more consistent results. The rest of the hyperparameters are set to the values in Table~\ref{tab:hyperparameters}.

We show our results for $n = 3, 4, 5, 6$ in Figure~\ref{fig:examples}, together with the analytical solution. The loss function for all of them is below $\num{2e-2}$.

\subsection{Wave equation\label{sec:wave}}

The wave equation is
\begin{equation}
  \frac{\partial^2 \phi}{\partial x^2} - \frac{\partial^2 \phi}{\partial t^2} = 0.
\end{equation}
For the initial and boundary conditions, we pick
\begin{gather}
  \phi(x, 0) = \sin(2 \pi x), \qquad
  \left.\frac{\partial\phi}{\partial t}\right|_{t = 0} = \pi \cos(2 \pi x),
  \\
  \phi(0, t) = \phi(1, t) = \sin(2 \pi t) / 2.
  \
\end{gather}
We use the natural choice of basis for the solution of this problem, which is the Fourier basis. Since the input space is 2-dimensional ($n_{\text{in}} = 2$), the basis is given as all products $\phi_{m_1}(x) \phi_{m_2}(t)$, with $\phi_0(z) = 1$, $\phi_1(z) = \sin(2 \pi z)$, etc. We set the number of $\phi_{m_i}$ functions per input component to $d = 3$, so the total dimension of the basis is 9. In order to reduce the number of qubits required for the encoding, we pick $n_{\text{spins}} = 2$, with the rest of hyperparameters set to the values in Table~\ref{tab:hyperparameters}.

We obtain a solution with a loss of $L = \num{3e-3}$. We present it in Figure~\ref{fig:examples}, together with the analytical solution, which is given by
\begin{multline}
  \phi_{\text{true}}(x, t) = \cos(2\pi x) \sin(2\pi t) \\
  + \sin(2\pi x) \cos(2\pi t) / 2.
\end{multline}

\subsection{Coupled first-order equations\label{sec:coupled}}

As an example of a system of coupled equations, we solve
\begin{gather}
  2 x' + x + 3 y = 0, \quad 2 y' + 3 x + y = 0, \\
  x(0) = 5, \quad y(0) = 3.
\end{gather}
In the $t \in [0, 1]$ interval. We use a monomial basis $x(t) = w_{0m} t^m$, $y(t) = w_{1m} t^m$. We pick the value $s_{nm} = 4$ for the scales of the binary encoding of the $w_{nm}$ weights, to account for the relatively large values of $x$ and $y$. We set all the other hyperparameters to the values in Table~\ref{tab:hyperparameters}. Finally, we find that we need to increase their relative importance in the loss function for the initial conditions to be satisfied. We do so by multiplying the equations (and not the initial conditions) by a factor of $1/10$.

We obtain a solution with a value of the loss of $\num{2e-2}$, and present it in Figure~\ref{fig:examples}, together with the analytical solution:
\begin{equation}
  x_{\text{true}}(t) = e^t + 4 e^{-2 t}, \quad y_{\text{true}}(t) = -e^t + 4 e^{-2 t}.
\end{equation}

\section{Conclusions\label{sec:conclusions}}

We have presented Qade, a general method for the solution of differential equations using quantum annealing devices. The first step is to re-formulate the equations as an optimization problem, by means of a general procedure which was originally developed for the application of machine learning. Then, this is transformed into a binary quadratic problem, which can directly be solved in a quantum annealer. The advantage of the quantum optimization is that it can tunnel through barriers, escaping local minima in which a classical optimizer might be trapped.

We have implemented Qade the proposed method in a Python package, described in Appendix~\ref{sec:qade} which provides a user-friendly interface for the calculation of the binary quadratic model corresponding to a set of equations, and for its solution in a D-Wave quantum annealer.

We have applied the method to 3 examples of differential equations, with features including variable coefficients, partial derivatives, and coupled equations. The current quantum annealing technology only allows to solve problems that require a low number of qubits to encode, but we find that the chosen equations can already be solved reliably. Thus, future quantum annealers with a larger number of qubits and a higher degree of connectivity have the potential to surpass classical methods for the solution of the larger differential equations that arise in real-world applications.

\appendix

\section{The \texttt{qade} Python package\label{sec:qade}}

We provide an implementation of the Qade method, described in Section~\ref{sec:method}, in the form of the Python package \texttt{qade}, which is publicly available in GitLab (\href{https://gitlab.com/jccriado/qade}{gitlab.com/jccriado/qade}) and PyPI, from which it can be installed through:
\begin{lstlisting}{bash}
 > pip install qade
\end{lstlisting}
For \texttt{qade} to send problems to be solved in the D-Wave systems, an installation of the D-Wave Ocean Tools, with the access token configured, is required. If this is not present, \texttt{qade} can still be used to compute the Ising model, whose ground state represents the solution to a given set of equations. In the rest of this Section, we describe \texttt{qade}'s interface in full generality. For examples of use, see~\href{https://gitlab.com/jccriado/qade/-/tree/main/examples}{gitlab.com/jccriado/qade/-/tree/main/examples}.

\subsection{Defining the problem}

The input data for \texttt{Qade} consists of the sets $X_i$ of sample points for the equations to be solved, together with the values of the vectors coefficients $C_{in}^{(k)}(x)$ and inhomogeneous terms $B_i(x)$ defined in Eq.~\eqref{eq:equations}, evaluated at all points $x \in X_i$. For easiness of use, an interface allowing for the specification of these parameters through a symbolic expression for an equation is provided. In order to define an equation
\begin{multline}
  c_1(x) \frac{\partial^k f_1}{\partial x_1^{k_1} \ldots \partial x_{n_{\text{in}}}^{k_{n_{\text{in}}}}}
  \, + \, c_2(x) \frac{\partial^l f_2}{\partial x_1^{l_1} \ldots \partial x_{n_{\text{in}}}^{l_{n_{\text{in}}}}} \\
  + \ldots
  + b(x) = 0,
\end{multline}
the user would write the code:
\begin{lstlisting}
f1, f2, ... = qade.function(n_in, n_out)
eq = qade.equation(
  c1 * f1[k1, ...] + c2 * f2[l1, ...]
  + ... + b,
  samples,
)
\end{lstlisting}
where \texttt{samples} is an array-like\footnote{An array-like object is either a \texttt{numpy} array or an object that can be converted into one. This includes scalars, lists and tuples.} with shape \texttt{(n\_samples, n\_in)} (or just \texttt{(n\_samples,)} when \texttt{n\_in == 1}), representing the set of samples $X_i$; while \texttt{c1}, \texttt{c2}, \ldots, \texttt{b} are either scalars or array-like objects of shape \texttt{(n\_samples,)}, giving the values of the corresponding parameters in the equation at all the sample points.

\subsection{Solving the problem}

The first step in solving a given set of equations is choosing an adequate basis of functions. The function
\begin{lstlisting}
basis = qade.basis(
  name, size_per_dim, n_in=1, scale=1.0
)
\end{lstlisting}
provides access to 5 pre-defined bases, which are listed in Table~\ref{tab:bases}. The \texttt{size\_per\_dim} defines the $d$ parameter for the first 3 bases in the table, and how many grid points per input-space dimension are defined for the last 2. In both cases, the total dimension of the bases is \texttt{size\_per\_grid ** n\_in}. The last two arguments, \texttt{n\_in} and \texttt{scale} are only used by the last 2 bases. \texttt{scale} corresponds to $\lambda$ in the definition of the corresponding $\phi(r)$ functions.

\begin{table*}
  \centering
  \begin{tabular}{ccc}
    \toprule
    Basis functions
    & Name
    & Definition \\
    \midrule
    \\[-5pt]
    \multirow{5}*{
    $\begin{array}{c}
       \Phi_m(x) = \phi_{m_1}(x_1) \phi_{m_2}(x_2) \ldots \\[5pt]
       (0 \leq m_i < d, \quad m = m_1 + m_2 d + \ldots)
     \end{array}$
    }
    & \texttt{"fourier"}
    & $\phi_m(x) = \left\{\begin{array}{ll}\cos(\pi n x) \\ \sin(\pi (n + 1) x)
                          \end{array}\right.$
    \\ \\
    & \texttt{"monomial"}
    & $\phi_m(x) = x^m$
    \\ \\
    & \texttt{"trig"}
    & $\phi_m(x) = \cos^{d - m - 1}(x) \sin^m(x)$ \\ \\
    \midrule
    \\[-5pt]
    \multirow{3}*{
    $\begin{array}{c}
       \Phi_m(x) = \phi(|x - z_m|) \\[5pt]
       (z_m \in \text{equally-spaced grid in } {[0, 1]}^{n_{\text{in}}})
     \end{array}$
    }
    & \texttt{"gaussian"}
    & $\phi(r) = -\exp\left[- (r / \lambda)^2\right]$
    \\ \\
    & \texttt{"multiquadric"}
    & $\phi(r) = \sqrt{r^2 + \lambda^2}$
    \\[10pt]
    \bottomrule
  \end{tabular}
  \caption{Bases of functions implemented by \texttt{Qade}.}
  \label{tab:bases}
\end{table*}

Given a list of equations \texttt{equations = [eq1, eq2, ...]} and a basis, \texttt{Qade} computes the quadratic loss function using Eqs.~\eqref{eq:quad-loss-J} and~\eqref{eq:quad-loss-h}. As a simplification, the pairs of indices $nm$ and $pq$ are flattened as $N = n m_{\text{max}} + m$ and $P = p q_{\text{max}} + q$, so that $J$ and $h$ become a matrix $J_{PQ}$ and a vector $h_P$. They are obtained using the function
\begin{lstlisting}
J, h = qade.loss(equations, basis)
\end{lstlisting}

The corresponding parameters $\hat{J}$ and $\hat{h}$ for the Ising model Hamiltonian are computed using Eqs.~\eqref{eq:ising-J} and~\eqref{eq:ising-h}. They are also flattened into a matrix $\hat{J}_{\hat{N}\hat{P}}$ and a vector $\hat{h}_{\hat{N}}$, through $\hat{N} = \alpha N_{\text{max}} + N$ and $\hat{P} = \beta P_{\text{max}} + P$, so that they can be directly provided to the D-Wave framework to be embedded in a quantum annealer. They are given by the function:
\begin{lstlisting}
J_hat, h_hat = qade.ising(
  equations, basis,
  n_spins, centers, scales,
)
\end{lstlisting}
The last three arguments are optional. \texttt{n\_spins} (default: 3) is the number of spins to use per weight $w_N$. \texttt{centers} (default: array of zeros) and \texttt{scales} (default: array of ones) are the flattened arrays of center values $c_N$ and scales $s_N$ from the binary encoding in Eq.~\eqref{eq:binary-encoding}.

The complete process of finding these parameters, sending them to a D-Wave QPU, setting it up and running the annealing process, reading the results, and decoding them back into the $w_{nm}$ matrix of weight is automatized by a single function call:
\begin{lstlisting}
sol = qade.solve(equations, basis, ...)
\end{lstlisting}
The returned object \texttt{sol} is a callable that receives an array \texttt{x} of samples and returns the value \texttt{sol(x)} of the solution at the sample points. It also contains 3 attributes:
\begin{itemize}
  \item \texttt{sol.basis}, the basis of functions in which the problem was solved (its name is available through \texttt{sol.basis.name}).
  \item \texttt{sol.weights}, the matrix $w_{nm}$ of weights.
  \item \texttt{sol.loss}, the value of the loss function.
\end{itemize}

\bibliographystyle{inspire}
\bibliography{references}

\end{document}